\documentclass[twocolumn,showpacs,showkeys,aps]{revtex4}
\usepackage{graphicx}
\usepackage{dcolumn}
\usepackage{bm}
\begin{document}
\newcommand{\ti}[1]{\mbox{\tiny{#1}}}
\newcommand{\im}{\mathop{\mathrm{Im}}}
\def\be{\begin{equation}}
\def\ee{\end{equation}}
\def\bea{\begin{eqnarray}}
\def\eea{\end{eqnarray}}

\title{A stationary $q-$metric}

\author{Saken Toktarbay$^1$ and Hernando Quevedo$^{2,3}$ }
\affiliation{
$^1$Department of Physics, Al-Farabi Kazakh National University, 050040 Almaty, Kazakhstan\\
$^2$Instituto de Ciencias Nucleares, Universidad Nacional Aut\'onoma de M\'exico, AP 70543, M\'exico, DF 04510, Mexico\\
$^3$Dipartimento di Fisica and ICRA, Universit\`a di Roma "La Sapienza", I-00185 Roma, Italy\\
}
\email{quevedo@nucleares.unam.mx,saken.yan@yandex.com}

\date{\today}

\begin{abstract}
We present a stationary generalization of the static $q-$metric, the simplest generalization of the Schwarzschild solution that contains a quadrupole parameter. It possesses three independent parameters that are related to the mass, quadrupole moment and angular momentum. We investigate the geometric and physical properties of this exact solution of Einstein's vacuum equations, and show that it can be used to describe the exterior gravitational field of rotating, axially symmetric, compact objects.

\end{abstract} 
\pacs{04.20.Jb;95.30.Sf}
\keywords{Quadrupole moment, q-metric, compact objects}

\maketitle

\section{Introduction}
\label{sec:int}

The black hole uniqueness theorems \cite{heussler}  state that the most general asymptotically flat solution of Einstein's vacuum field equations with a regular horizon is the Kerr metric that possesses two independent parameters only, corresponding to the mass and angular momentum. In terms of multipole moments, this statement is equivalent to saying that black holes can have only mass monopole and angular dipole moments. All the higher multipole moments must disappear, probably in the form of gravitational waves, during the gravitational collapse of an arbitrary rotating mass distribution whose final state is a black hole.

On the other hand, astrophysical compact objects include not only black holes, but also regular stars, neutron stars, white dwarfs, planets, etc. For the description of the gravitational field of such objects, one can expect that higher multipole  moments could play an important role. Let us consider the particular case of a static mass distribution with a quadrupole moment that describes the deviation from spherical symmetry. The uniqueness theorems demonstrate that, in the case of vanishing quadrupole, there exists only one vacuum solution, namely, the Schwarzschild solution. As soon as a non-vanishing quadrupole is considered, the uniqueness is no more valid and so there could be in principle an infinite number of vacuum solutions with mass and quadrupole parameters. The first vacuum solution with a quadrupole parameter was derived by Weyl in 1917 \cite{weyl17}. Today,  many other solutions are known, including their stationary generalizations \cite{er59,dh82,islam1985,manko1992,manko1990a,manko1990b,manko2000,prs06,qm91,solutions}. Many other static solutions can be generated by using the fact that the field equations are linear and by applying certain differential operators to the harmonic \cite{quev87}. One common problem of all these solutions is that they are difficult to be handled due to their complicated structure. Recently, it was proposed to reinterpret the Zipoy-Voorhees metric \cite{zip66,voor70} as a generalization of the Shwarzschild metric with a quadrupole parameter $(q-$metric). To our knowledge, the $q-$metric is the simplest static generalization of the Schwarzschild spacetime with an additional parameter that determines an independent mass quadrupole moment. 

The aim of this paper is to derive a stationary generalization of the $q-$metric to take into account the rotation of the quadrupolar mass distribution. We will show that this generalization satisfies all physical conditions to be considered as a candidate to describe the exterior gravitational field of deformed compact objects.

This paper is organized as follows. In Section \ref{sec:qmet}, we review the main properties of the $q-$metric. In Section \ref{sec:qstar}, we present the corresponding stationary generalization and analyze its main physical properties. Section \ref{sec:con} is devoted to the conclusions.

\section{The $q-$metric and its properties}
\label{sec:qmet}

The $q-$metric in spherical coordinates can be expressed as \cite{quev11}
\bea
ds^2 = && h^{1+q} dt^2  
- h^{-q}\bigg[ \left(1+\frac{m^2\sin^2\theta}{r^2 h}\right)^{-q(2+q)}\nonumber\\
&& \times\left(\frac{dr^2}{h}+ r^2d\theta^2\right) + r^2 \sin^2\theta d\varphi^2\bigg],\\
&& h= 1-\frac{2m}{r}\nonumber\ .
\label{qmet}
\eea 
This is an asymptotically flat vacuum solution of Einstein's equation. 
The physical interpretation of the parameters $m$ and $q$ can be clarified by calculating 
the invariant Geroch multipoles 
\cite{ger}:
\be 
M_0= (1+q)m\ , \quad M_2 = -\frac{m^3}{3}q(1+q)(2+q)\ .
\ee
Higher moments are proportional to $mq$ and can be  completely rewritten in terms of $M_0$ and $M_2$; 
accordingly, the parameters $m$ and $q$ determine the mass and quadrupole 
In the limiting case $q=0$ only the monopole $M_0=m$ 
survives, as in the Schwarzschild spacetime. In the limits $m=0$ with arbitrary $q$ and $q=-1$ with arbitrary $m$, all moments vanish identically and the spacetime becomes flat.  
The deviation from spherical symmetry is described by the quadrupole moment $M_2$ which is positive for  prolate sources 
and negative for  oblate sources. Since the total mass $M_0$ must be positive, we have  that $q>-1$ (we assume $m>0$). 

An investigation of the Kretschmann scalar shows that the hypersurface $r=2m$ is always singular for any non-vanishing value of $q$. Moreover, $r=0$ is also a singularity. Depending on the value of $q$, additional singularities can appear that are always situated inside the exterior singularity located at $r=2m$. 

All these properties indicate that the $q-$metric can be used to describe the exterior gravitational field of deformed mass distribution. It also describes the field of a naked singularity situated at $r=2m$. From the physical point of view, this is not a problem because one can ``cover" the naked singularity with an interior solution that must be matched with the exterior $q-$metric at some radius $r_{matching} > 2m$. 

To present the stationary generalization of the $q-$metric, it is convenient to introduce prolate spheroidal coordinates given by 
\be
x=\frac{r}{m} - 1\ ,\quad y=\cos\theta\ .
\ee

\section{The rotating $q-$metric}
\label{sec:qstar}

The general stationary axisymmetric line element in prolate spheroidal coordinates  is given as 
\bea
& & ds^2 = f(dt-\omega d\varphi)^2 - \frac{\sigma^2}{f}\bigg[ e^{2\gamma}(x^2-y^2) \nonumber \\
& &
\times \left( \frac{dx^2}{x^2-1} + \frac{dy^2}{1-y^2} \right) 
+ (x^2-1)(1-y^2) d\varphi^2\bigg] \ ,
\label{lelxystar}
\eea
where $\sigma=$const. and all the metric functions depend on $x$ and $y$ only.

It turns out to be useful to introduce the 
the complex Ernst potential \cite{ernst}
\be
E=f+i\Omega \ , 
\label{ernst1}
\ee
where the function $\Omega$ is now determined by the equations
\be
 \sigma (x^2-1)\Omega_x = f^2\omega_y\ ,\quad  \sigma (1-y^2)\Omega_y = - f^2 \omega_x \ . 
\label{ernst2}
\ee
We can see that if the potential $E$ is given, the metric function $f$ can be found algebraically and the metric function $\omega$ is computed by quadratures from Eqs.(\ref{ernst2}). Moreover, the metric function $\gamma$ is determined by two first-order differential equations that can be integrated by quadratures once $E$ is known. It follows that all the information about the metric (\ref{lelxystar}) is contained in the Ernst potential $E$. 

To obtain the explicit form of the new Ernst potential, we use the solution generating techniques \cite{dietz} that allow us to generate stationary solutions from a static solution. If we take as seed solution the $q-$metric in prolate spheroidal coordinates, several differential equations must be solved to obtain the explicit form of the Ernst potential. The details of this derivation will be presented elsewhere. The final expression for the Ernst potential can be written as
\be
E= \left(\frac{x-1}{x+1}\right)^q \frac{x-1+(x^2-1)^{-q} d_+} {x+1+ (x^2-1)^{-q} d_- }\ ,
\label{qstar1}
\ee 
with
\be
d_\pm = \alpha^2 (1\pm x)h_+ h_- + i \alpha [y (h_++h_-)\pm (h_+-h_-)]\ ,
\ee
\be
h_\pm = (x\pm y)^{2q} \ .
\ee
Here, we have a new arbitrary parameter $\alpha$. As expected, in the limiting case $\alpha=0$, we obtain the $q-$metric. 

By analyzing the behavior of the Ernst potential, one can prove that this new solution is asymptotically flat. The computation of the corresponding metric functions corroborates this result. Moreover, the behavior at the axis $y=\pm 1$ shows that it is free of singularities outside a region which is always inside the radius $x_s=\frac{m}{\sigma}$, which in the case of vanishing $\alpha$ corresponds to the exterior singularity situated at $r_s=2m$. The expression for the Kretschmann scalar is quite cumbersome and cannot be written in a compact form. Nevertheless, a careful numerical and analytical inspection shows that the outer most singularity is situated at $x_s=\frac{m}{\sigma}$. Inside this radius, several singular structures can appear that depend on the value of $q$ and $\sigma$. 

To find out the physical meaning of the parameters entering the new metric, we compute the coordinate invariant multipole moments as defined by Geroch \cite{ger}, using a procedure proposed in \cite{quev90} that allows to perform the computations directly from the Ernst potential. First, let us consider the particular limiting case with $q=0$. Then, choosing the new parameter as $\alpha=\frac{\sigma-m}{a}$ the  resulting multipoles are
\begin{equation} 
M_{2k+1} = J_{2k}=0 \ ,  \quad k = 0,1,2,... 
\end{equation} 
\begin{equation} 
M_0 = m \ , \quad M_2 = - ma^2   \ , ... 
\end{equation} 
\begin{equation} 
J_1= ma \ , \quad J_3 = -ma^3  \ , ....
\end{equation}
These are the mass $M_n$ and angular $J_n$ multipole moments of the Kerr solution. From the corresponding Ernst potential one can derive the Kerr metric in prolate spheroidal coordinates. This result shows that indeed the new solution given in Eq.(\ref{qstar1}) contains information about the rotation of the source. 

In the general case of arbitrary $q$ parameter, we obtain the following multipole moments 
\be
M_0= m+\sigma q\ ,
\ee
\be
M_2=7/3\,{\sigma}^{3}q-1/3\,{\sigma}^{3}{q}^{3}+m{\sigma}^{2}-m{\sigma}^{2
}{q}^{2}-3\,{m}^{2}\sigma\,q-{m}^{3}\ ,
\ee
\be
J_1= ma + 2a \sigma q\ ,
\ee
\bea
J_3 = && -1/3\,a ( -8\,{\sigma}^{3}q+2\,{\sigma}^{3}{q}^{3}-3\,m{\sigma}^{
2}+9\,m{\sigma}^{2}{q}^{2}\nonumber\\
&& +12\,{m}^{2}\sigma\,q+3\,{m}^{3})\ .
\eea
It can be shown that the even gravitomagnetic and the odd gravitoelectric multipoles vanish identically 
because the solution has an additional reflection 
symmetry with 
respect to the plane $y=0$ which represents the equatorial plane.  
Higher odd gravitomagnetic and even gravitoelectric multipoles can be shown to be linearly dependent since they are completely determined by the parameters $m$, $a$, $\sigma$ and $q$.

\section{Conclusions}
\label{sec:con}

In this work, we presented a stationary generalization of the static $q-$metric, which is the simplest generalization of the Schwarschild metric containing a quadrupole parameter. The new solution was given in terms of the Ernst potential from which all the metric functions can be derived algebraically or by quadratures.

The stationary $q-$metric turns out to be asymptotically flat and free of singularities outside a region determined by the spatial coordinate $x_s =\frac{m}{\sigma}$, which in the static limiting case is situated at the singular hypersurface $r_s=2m$. The new solution contains as a particular case the Kerr solution, indicating that the new free parameter can be associated with rotation of the mass distribution. We conclude that the stationary $q-$metric can be used to describe the exterior gravitational field of a rotating deformed mass distribution.

It can be expected that for realistic compact objects the exterior singularity is situated very closed to the center of the body, because in the static limit it is located at the Schwarzschild radius of the mass distribution. Therefore, it should be possible to cover the exterior singularity with a suitable interior solution. This is the task of future investigations.

\section*{Acknowledgements}

This work was  supported by DGAPA-UNAM, Grant No. 106110, and Conacyt-Mexico, Grant No. 166391.  


\end{document}